\documentclass{article}
\usepackage{amssymb}
\usepackage{amsmath}
\usepackage{graphicx}

\begin{document}

\begin{center}
{\large \textbf{Note on the Schwarzschild-phantom wormhole}}

\bigskip

\textbf{R Lukmanova$^{1,*}$, A Khaibullina$^{1}$, R Izmailov$^{1}$, A Yanbekov$^{1}$,}

\textbf{R. Karimov$^{1}$ and A A Potapov$^{2}$}

\bigskip 

$^{1}$Zel'dovich International Center for Astrophysics, M. Akmullah Bashkir
State Pedagogical University, Ufa 450000, RB, Russia

$^{2}$Department of Physics and Astronomy, Bashkir State University,
Sterlitamak Campus, Sterlitamak 453103, RB, Russia \\[0pt]

\bigskip

\end{center}

\textbf{Abstract:} Recently, it has been shown by Lobo, Parsaei and Riazi (LPR) that phantom
energy with $\omega =p_{r}/\rho <-1$ could support phantom wormholes.
Several classes of such solutions have been derived by them. While the inner
spacetime is represented by asymptotically flat phantom wormhole that have
repulsive gravity, it is most likely to be unstable to perturbations. Hence,
we consider a situation, where a phantom wormhole is somehow trapped inside
a Schwarzschild sphere across a thin shell. Applying the method developed by
Garcia, Lobo and Visser (GLV), we shall exemplify that the shell can possess
zones of stability depending on certain constraints. It turns out that zones
corresponding to "force" constraint are more restrictive than those from the
"mass" constraint. We shall also enumerate the interior energy content by
using the gravitational energy integral proposed by Lynden-Bell, Katz and Bi%
\v{c}\'{a}k. It turns out that, even though the interior mass is positive,
the integral implies repulsive energy. This is consistent with the phantom
nature of interior matter.

\bigskip


\bigskip

\textbf{PACS No(s)}: 04.20.Jb, 04.20.Gz, 98.80.Es

\bigskip
\textbf{1. Introduction}
\bigskip

Wormholes are topological tunnels that connect two universes or two distant
regions of spacetime. The subject is more of an academic curiosity but
nonetheless has received some attention after the influential work by Morris
and Thorne [1]. Wormholes have not been observed in any experiment but have
not yet been ruled out by observations either. Again, there have been no
known experiments specifically designed to observe or rule out such objects.
On the other hand, there is enormous work on black holes, especially
directed to finding a supermassive black hole at our galactic center%
\footnote{%
We thank an anonymous reviewer for pointing it out.}. There could
nonetheless be a modest interest in the topic of wormholes so long as the
exterior does not deviate from the Schwarzschild vacuum but allows other
matter like phantom in the interior. We then call it phantom wormhole. Since
the well known regular Ellis wormhole is unstable [3,4], we look for stable
solutions, and hence study the stability of phantom wormholes.

Phantom wormholes could be interesting in the sense that they are built out
of phantom energy defined by equation of state $\omega =p_{r}/\rho <-1,$
which is speculated to be a possible cause driving late-time cosmic
acceleration [5]. Both wormholes or phantom equation of state have really
not been confirmed by observations. However, mathematical solutions exist.
It has been recently shown by Lobo, Parsaei and Riazi (LPR) that wormholes
with phantom energy could actually be found as exact solutions (as opposed
to artificially engineered ones) of Einstein's equations. Several classes of
such solutions have been derived by them. However, since the phantom
wormholes are built of the Null Energy Condition (NEC) violating exotic
matter (since $p_{r}+\rho <0$), they are likely to be unstable though an
exact formulation of such instability is still unavailable. We try to
imagine a situation, where some kind of stability involving phantom
wormholes would still be possible.

To that end, we thought it useful to consider that a phantom wormhole is
somehow trapped inside a Schwarzschild sphere across a thin shell. We shall
employ a novel and recent formalism to study the stability of such thin
shells developed by Garcia, Lobo and Visser (GLV) [6]. We note that the
stability of wormholes with phantom energy has been analyzed in [7], but the
author consider other configurations of wormholes and took into account only
the "mass constraint". At the same time, we know that there is also another
constraint, viz., the "force constraint" and is more restrictive in nature,
as will be seen later, which allows us to obtain more precise ranges of
stability. The formalism was primarily used in Ref.[8] for the stability of
the thin-shell Schwarzchild-Ellis wormhole. We shall consider a particular
class of asymptotically flat LPR phantom wormholes, viz., that of bounded
mass gluing it to a vacuum Schwarzschild exterior. The configuration would
resemble a gravastar of certain radius having in its interior repulsive
phantom matter and a Schwarzschild vacuum in the exterior. For an outside
observer, there would therefore be no distinction between a true
Schwarzschild mass and a gravastar of this kind.

In the present paper, we shall extend the stability analysis of the
Schwarzschild - phantom thin-shell wormhole to include also the "external
force" term [6], and delineate the zones of stability. In addition, we shall
calculate the interior energy content by using the gravitational energy
integral proposed by Lynden-Bell, Katz and Bi\v{c}\'{a}k [10,11]. It turns
out that, even though the interior mass in positive, the integral implies
repulsive energy in the interior. This is consistent with the fact that the
interior mass is phantom in nature.

The paper is organized as follows: In Sec.2, we shall outline the GLV
method. In Sec.3, we outline the features of the LPR phantom wormhole under
consideration and in Sec.4, we enumerate the interior energy constant. In
Sec.5, we build the composite object gluing together at some radius the
phantom wormhole and the exterior Schwarzschild spacetime. Results are
discussed in Sec.6. We take units such that $G=c=1$, unless specifically
restored.

\bigskip
\textbf{2. The GLV method}
\bigskip

We shall only cite their results that will be used here and take the liberty
to restate the new concepts they have developed. For details of their
ingenious arguments and calculations, the reader is asked to read the
original paper [6]. They take the spacetimes on two $\pm $ sides of the thin
shell as

\begin{equation}
\ \ \ \ \ \ \ ds^{2}=-e^{2\Phi _{\pm }(r_{\pm })}\left[ 1-\frac{b_{\pm
}(r_{\pm })}{r_{\pm }}\right] dt_{\pm }^{2}+\left[ 1-\frac{b_{\pm }(r_{\pm })%
}{r_{\pm }}\right] ^{-1}dr_{\pm }^{2}+r_{\pm }^{2}d\Omega _{\pm }^{2}.
\end{equation}

The method allows any two arbitrary spherically symmetric spacetimes to be
glued together by cut and paste procedure. Thus, for the static and
spherically symmetric spacetime, the single manifold $\mathcal{M}$ is
obtained by gluing two bulk spherically symmetric spacetimes $\mathcal{M}%
_{+} $ and $\mathcal{M}_{-}$ at a timelike junction surface $\Sigma $, i.e.,
at $f(r,\tau )=r-a(\tau )=0$. The surface stress-energy tensor may be
written in terms of the surface energy density $\sigma $ and the surface
pressure $P$ as $S_{ij}=$diag$(-\sigma ,P,P)$. GLV work out the general
conservation law

\begin{equation}
\frac{d(\sigma A)}{d\tau }+P\frac{dA}{d\tau }=\Xi A \dot{a},
\end{equation}%
where $\dot{a}=\frac{da}{d\tau }$, the shell surface area $A=4\pi a^{2}$ and
there is an entirely new term 
\begin{equation}
\Xi =\frac{1}{4\pi a}\left[ \Phi _{+}^{\prime }(a)\sqrt{1-\frac{b_{+}(a)}{a}+\dot{a}^{2}}+\Phi _{-}^{\prime }(a)\sqrt{1-\frac{b_{-}(a)}{a}+\dot{a}^{2}}%
\right].
\end{equation}%
The first term in Eq.(2) represents the variation of the internal energy of
the shell, the second term is the work done by the internal force of the
shell. The right hand side is the net discontinuity in the conservation law
of the surface stresses of the bulk momentum flux and is physically
interpreted as the work done by external forces on the thin shell. In short
it is the "external force" term occurring due to $\Phi _{\pm }^{\prime }\neq
0$. This is a new concept not noted so far in the literature. Only when $%
\Phi _{\pm }^{\prime }(a)=0$, we have $\Xi =0$, and then one recovers the
familiar conservation law on the shell.

Assuming integrability of $\sigma $ [6], which allows $\sigma =\sigma (a)$,
it is possible to define the mass of the thin shell of exotic matter
residing on wormhole throat as [6] 
\begin{equation}
m_{s}(a)=4\pi \sigma (a)a^{2}.
\end{equation}%

GLV derived the workable master inequalities about stability around a given
radius $a_{0}$ after long calculations, which are the constraint from the
"mass"

\begin{center}
\begin{eqnarray}
m_{s}^{\prime \prime }(a_{0})&\geq \frac{1}{4 a_{0}^3} \left[\frac{%
[b_{+}(a_{0})-a_{0}b_{+}^{\prime }(a_{0})]^2}{[1-b_{+}(a_{0})/a_{0}]^{3/2}}+ \frac{[b_{-}(a_{0})-a_{0}b_{-}^{\prime }(a_{0})]^2}{[1-b_{-}(a_{0})/a_{0}]^{3/2}}\right] &  \notag \\
&+\frac{1}{2}\left[\frac{b_{+}^{\prime \prime }(a_{0})}{\sqrt{1-b_{+}(a_{0})/a_{0}}}+ \frac{b_{-}^{\prime \prime }(a_{0})}{\sqrt{1-b_{-}(a_{0})/a_{0}}}\right]
\end{eqnarray}
\end{center}

and when $\Phi _{\pm }^{\prime }(a_{0})\leq 0$, the constraint from the
"external force"

\begin{center}
\begin{eqnarray}
\left[ 4\pi \Xi (a)a\right] ^{\prime \prime }|_{a_{0}} &\geq &\left[ \Phi
_{+}^{\prime \prime \prime }(a)\sqrt{1-b_{+}(a)/a}+\Phi _{-}^{\prime
\prime \prime }(a)\sqrt{1-b_{-}(a)/a}\right] \big|_{a_{0}} 
 \notag \\
&&-\left[ \Phi _{+}^{\prime \prime }(a)\frac{(b_{+}(a)/a)^{\prime }}{\sqrt{1-%
b_{+}(a)/a}}+\Phi _{-}^{\prime \prime }(a)\frac{(b_{-}(a)/a)^{\prime
}}{\sqrt{1-b_{-}(a)/a}}\right] \big|_{a_{0}}  
\notag \\
&&-\frac{1}{4}\left[ \Phi _{+}^{\prime }(a)\frac{[(b_{+}(a)/a)^{\prime} ]^{2}}{%
[1-b_{+}(a)/a]^{3/2}}+\Phi _{-}^{\prime }(a)\frac{%
[(b_{-}(a)/a)^{\prime }] ^{2}}{[1-b_{-}(a)/a]^{3/2}}\right] \big|_{a_{0}}
\notag \\
&&-\frac{1}{2}\left[ \Phi _{+}^{\prime }(a)\frac{(b_{+}(a)/a)^{\prime \prime
}}{\sqrt{1-b_{+}(a)/a}}+\Phi _{-}^{\prime }(a)\frac{%
(b_{-}(a)/a)^{\prime \prime }}{\sqrt{1-b_{-}(a)/a}}\right] \big|%
_{a_{0}}.
\end{eqnarray}
\end{center}

Similar, but not the same, force constraint appears for $\Phi _{\pm
}^{\prime }(a_{0})\geq 0$ as well. We shall not quote it here as our example
has $\Phi _{\pm }^{\prime }(a_{0})\leq 0$. Such a force constraint however
disappears if $\Phi _{\pm }^{\prime }(a_{0})= 0$.

Eqs.(5) and (6) are the ones that will be used in the case of the LPR
wormhole.

\bigskip
\textbf{3. The LPR phantom wormhole}
\bigskip

We shall consider here only the solution with bounded mass. To be consistent
with notation with the above section, we shall change the constant in the
LPR wormhole [9] such that it reads%
\begin{equation}
ds^{2}=-\left[ 1+\frac{\lambda r_{0}}{r}\right] ^{1-\frac{1}{\lambda }%
}dt^{2}+\frac{dr^{2}}{1-\frac{r_{0}}{r}(\frac{\lambda r_{0}}{r}+1-\lambda )}%
+r^{2}d\Omega ^{2},
\end{equation}%
where $\lambda $ is a constant and $r_{0}$ is the minimum radius. According
to equation of state $\lambda \omega =-1$, we get the following range $%
-1<\lambda <0$. The mass function is%
\begin{equation}
m(r)=\frac{\lambda r_{0}}{2}\left( \frac{r_{0}}{r}-1\right) .
\end{equation}%

The throat appears at 
\begin{equation}
r_{\text{th}}=r_{0}\text{ and }\lambda r_{0},
\end{equation}%
but we ignore the second radius because it is negative. Thus the wormhole
spacetime is defined for $r_{0}<r<\infty $. The shape function and the
redshift function respectively are

\begin{equation}
b_{-}=r_{0}(\frac{\lambda r_{0}}{r}+1-\lambda),\qquad \Phi _{-}=\frac{1}{2}%
\ln \left[ \frac{\left[ 1+\frac{\lambda r_{0}}{r}\right] ^{1-\frac{1}{\lambda%
}}}{1-\frac{r_{0}}{r}(\frac{\lambda r_{0}}{r}+1-\lambda)}\right] .
\end{equation}%

Using the Einstein field equation, $G_{\mu \nu }=8\pi T_{\mu \nu }$, the
(orthonormal) stress-energy tensor components in the bulk are given by

\begin{equation}
\rho =-\frac{\lambda r_{0}^2}{8\pi r^{4}},
\end{equation}

\begin{equation}
p_{r}=-\frac{r_{0}^{2}}{8\pi r^{4}},
\end{equation}

\begin{equation}
p_{t}=\frac{r_{0}^{2}(4r+(\lambda^{2}+4\lambda-1)r_{0})}{32\pi
r^{4}(r+\lambda r_{0})}.
\end{equation}%

Here $\rho $ is the energy density, $p_{r}$ is the radial pressure, and $%
p_{t}$ is the lateral pressure measured in the orthogonal direction to the
radial direction. We thus obtain the NEC violating condition 
\begin{equation}
\rho +p_{r}=-\frac{(1+\lambda)r_{0}^{2}}{8\pi r^{4}}<0,\text{ }\forall r.
\end{equation}

\bigskip
\textbf{4. Energetics in the phantom wormhole}
\bigskip

We shall use this form of the metric for gluing with the Schwarzschild black
hole exterior and place the mass in the interior of the thin shell. Using
(7), one has 
\begin{equation}
\rho =\frac{1}{8\pi r^{2}}\frac{db}{dr}=-\frac{\lambda r_{0}^{2}}{8\pi r^{4}}%
>0,\qquad \Omega _{\text{WEC}}=\frac{1}{2}\int\limits_{r_{0}}^{\infty }\rho
r^{2}dr=-\frac{\lambda r_{0}}{16\pi }>0,
\end{equation}%
where $\Omega _{\text{WEC}}$ is the total amount of Weak Energy Condition
(WEC) violating matter, which, in this case, is positive. To understand the
nature of gravity (whether attractive or repulsive) due to the interior mass
defined by Eq.(8), let us consider the Lynden-Bell, Katz and Bi\v{c}\'{a}k
[10,11] form of gravitational energy $E_{G}$ defined by 
\begin{equation}
E_{G}=\Omega _{\text{WEC}}-E_{M},
\end{equation}%
where $E_{M}$ is the sum of other forms of energy like rest energy, kinetic
energy, internal energy etc. and is defined by 
\begin{equation}
E_{M}=\frac{1}{2}\int\limits_{r_{0}}^{\infty }\rho (g_{rr})^{1/2}r^{2}dr.
\end{equation}%

In the present case, we find that 
\begin{eqnarray}
\;E_{M} &=&\frac{\sqrt{\lambda }r_{0}\left( \pi -2\text{ }\arcsin \left[ 
\frac{1}{\sqrt{1+\lambda }}\right] \right) }{16\pi }<0  \notag \\
&\Rightarrow &E_{G}=-\frac{\sqrt{\lambda }r_{0}\left( \sqrt{\lambda }+\pi
-2\arcsin \left[ \frac{1}{\sqrt{1+\lambda }}\right] \right) }{16\pi }>0,
\end{eqnarray}%
which implies that, even though the mass in Eq.(8) is positive, the
gravitational energy $E_{G}$ of the interior matter is positive, hence 
\textit{repulsive} in nature (Fig.1). This only points to the repulsive
nature of phantom energy. This interior repulsion is balanced by the
exterior attraction due to the Schwarzschild mass at a place where the
phantom wormhole's surface is formed. Similar constructions exist in the
literature [12-17]. For example, the static spherically symmetric vacuum
condensate star (gravastar) devised by Mazur and Mottola [16] has an
isotropic de-Sitter vacuum in the interior, the matter marginally satisfying
the NEC and strictly violating the Strong Energy Condition (SEC). It has a
Schwarzschild exterior $(p=0,\rho =0)$ of mass $M$. The difference with our
case is that we have a phantom wormhole interior instead of the de-Sitter
vacuum.

\begin{figure}[tbp]
\centerline { 
\includegraphics [scale=0.6]{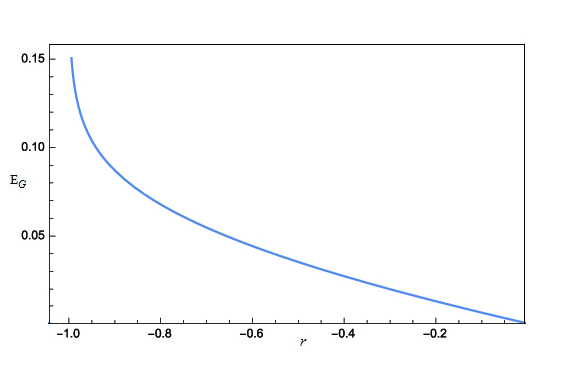}}
\caption{Energetics of the LPR wormhole, where $\protect\lambda\in\left[-1,0%
\right]$}
\end{figure}

\bigskip
\textbf{5. Schwarzschild - phantom wormhole }
\bigskip

We glue the horizonless LPR phantom wormhole with Schwarzschild exterior at
some given radius $r=a_{0}>2M$ , $r_{0}$ (throat radius of phantom
wormhole), that is, we just join the two spacetimes, LPR and Schwarzschild,
at a radius above the Schwarzschild horizon, i.e., at $r>r_{\text{hor}}=2M$.
The interior regions $r\leq 2M$, $r_{0}$ are surgically excised out from
respective spacetimes because we don't want the presence of any horizon in
the resultant wormhole. This $a_{0}$ is the radius about which linear
spherical perturbations are assumed to take place. Casting the Schwarzschild
metric in the GLV form (1), we get 
\begin{equation}
b_{+}=2M,\qquad \Phi _{+}=0
\end{equation}%
and similarly casting phantom wormhole metric (7) in the form of GLV metric
(1), we get Eqs.(10).

Since $-1<\lambda<0$, we have $\Phi _{\pm }^{\prime }\leq 0$ for $%
a_{0}>r_{0} $, and there will occur the effect of "external force"
influencing the thin shell motion. Putting the above functions in the
inequalities (5,6), and defining

\begin{equation}
x=\frac{M}{a_{0}},\qquad y=\frac{r_{0}}{M},
\end{equation}%
we find, respectively

\begin{center}
\begin{eqnarray}
4 a_{0} m_{s}^{\prime \prime }(a_{0}) \geq f(x,y) =& \frac{4}{(1-2 x)^{3/2}}+%
\frac{4 \lambda y^2}{\sqrt{(1-xy)(1+\lambda y x)}}  \notag \\
&+\frac{y^2(1-\lambda+2 \lambda y x)^2 }{((1-xy)(1+\lambda xy))^{3/2}} ,
\end{eqnarray}
\begin{eqnarray}
8a_{0}^{3}\left[ 4\pi \Xi (a)a\right] ^{\prime \prime }|_{a_{0}} &\geq
g(x,y)=\frac{-x^{2}(1+\lambda)}{((1-x)(1+\lambda x))^{5/2}}{\LARGE [}%
(48+12\lambda^{2}x^{4}  \notag \\
&+4x(\lambda-1)(19-10x^{2}\lambda+\sqrt{(1-x)(1+\lambda x)}))  \notag \\
&+x^{2}(31+31\lambda^{2}-2\lambda(55+6\sqrt{(1-x)(1+\lambda x)})){\LARGE ]}
\end{eqnarray}
\end{center}

\begin{figure}[tbp]
\centerline { \includegraphics 
[scale=0.7]{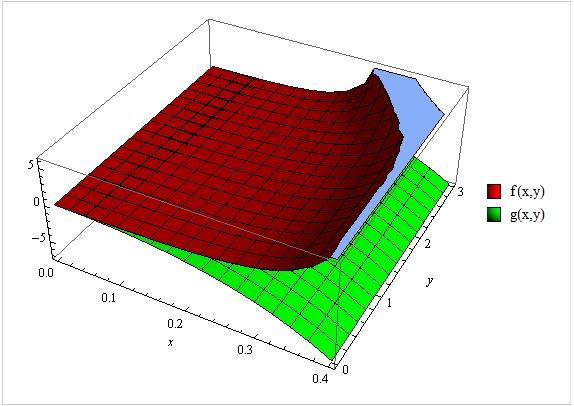}}
\caption{Stability zone for the Schwarzschild - Phantom wormhole with a
bounded mass function thin shell. The ranges are defined by $x=\frac{M}{a_{0}%
}$, $y=\frac{r_{0}}{M}$, where $x\in \left[ 0,0.4\right] $, $y\in \left[ 0,3%
\right] $ and $\protect\lambda =-0.5$.}
\end{figure}

\begin{figure}[tbp]
\centerline { 
\includegraphics
[scale=0.7]{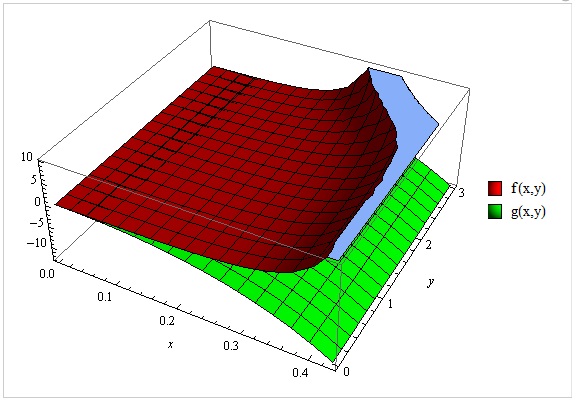}}
\caption{Stability zone for the Schwarzschild - Phantom wormhole with a
bounded mass function thin shell. The ranges are defined by $x=\frac{M}{a_{0}%
}$, $y=\frac{r_{0}}{M}$,, where $x\in \left[ 0,0.4\right] $, $y\in \left[ 0,3%
\right] $ and $\protect\lambda =-0.2$.}
\end{figure}

\bigskip
\textbf{6. Results and discussion }
\bigskip

A possible candidate for the accelerated expansion of the Universe is
speculated to be phantom energy, a cosmic fluid governed by an equation of
state of the form $\omega =p_{r}/\rho <-1$, which consequently violates the
null energy condition. We have analyzed the physical properties and
characteristics of thin-shell phantom wormholes gluing with exterior
Schwarzschild spacetime. The reason for choosing Schwarzschild exterior is
two fold: first, this exterior spacetime is the simplest and most well
discussed, especially in the context of the existence of a black hole in the
galactic center. Second, the speculation of phantom energy curling up by
some high energy process into a Schwarzschild-like star is by itself of
interest.

We have analyzed the stability regions of such configurations by including
the effects of "external forces", in addition to that of the "mass" term.
The "force constraint" is just a momentum flux across the shell and is
included here only for completeness. We enumerated the interior energy
content by using the gravitational energy integral by Lynden-Bell, Katz and
Bi\v{c}\'{a}k [10,11]. It turns out that, even though the interior mass in
positive, the integral is positive implying repulsive energy in the
interior. This is consistent with the fact that the mass is phantom in
nature. To make stability analyses physically meaningful, one should be able
to delineate the possible parameter ranges for which stability can be
achieved. The general and unified GLV stability analysis provides an
excellent way to achieve this goal. Our example above is interesting in the
sense that the outer mouth has a Schwarzschild mass ($M\neq 0$) as viewed by
observers in the first asymptotic region and the inner mouth has nonzero
asymptotic mass as viewed by observers in the second asymptotic region. So
the configuration resembles that of a gravastar.

The stability condition can be expected as an explicit inequality involving
the second derivative of the "mass" of the throat, $m_{s}^{\prime \prime
}(a) $. In the absence of "external forces", this inequality is given by
(5), and this is the only stability constraint one requires. However, once
one has external forces ($\Phi _{\pm }^{\prime }(a_{0})=0$), there arises
additional constraints in the form of inequality (6), which is imposed from
the external force over those from the linearized perturbations around
static solutions of the Einstein field equations, as worked out above.

Our results show the following: Fig.2 $f(x,y)$ refers to the effect on
stability of the motion of thin-shell due to its "mass" and $g(x,y)\;$--
effect due to the "external force" ($\lambda =-0.5$). The region above the
surface which is given by function $g(x,y)$ is the stable region. It can be
seen that, as $x$ increases, the stability region gradually diminishes to
zero near $x\sim \frac{1}{3}$. Thus the thin-shell motion becomes unstable
very near the Schwarzschild horizon but this phenomenon is insensitive to
values of $y$\ at any fixed$\;$value of $x$ . The region above the surface
which is given by function $f(x,y)$ is the stable region. We see that, when $%
x$ increases \ from $0$ to $\frac{1}{3}$, the stability region steadily
diminishes finally to zero near higher values of $x\sim \frac{1}{3}$.
However, as $y$ increases from $0$ to $3$, the region of stability steadily
decreases until it diminishes completely near the higher end, $y\sim 3$.

Thus the sensitivity in the $y$-direction, absent for $g(x,y)$, is revealed
for $f(x,y)$. Note that the "mass" constraint (5) does not involve $\Phi
_{\pm }$ and its derivatives, while the "force" constraint (6) involves them
and in fact exists due to them. Similar results are obtained with $\lambda
=-0.2$ and shown in Fig.3. Our analysis confirms that the force constraint
provide a more complete stability zone than those from the mass constraint.
This informative picture has been available thanks only to the effect of the
newly developed "external force" constraint by GLV.

\bigskip
\textbf{References}
\bigskip

[1] M S Morris and K S Thorne, Am. J. Phys. \textbf{56}, 395 (1988).

[2] J A Gonzalez, F S Guzman and O Sarbach, Class. Quant. Grav. \textbf{26},
015010 (2009).

[3] J A Gonzalez, F S Guzman and O Sarbach, Class. Quant. Grav. \textbf{26},
015011 (2009).

[4] K A Bronnikov, J C Fabris and A Zhidenko, Eur. Phys. J. C \textbf{71},
1791 (2011).

[5] G Hinshow et al., Astrophys. J. Suppl. Ser. \textbf{148}, 135 (2003)

[6] N M Garcia, F S N Lobo and M Visser, Phys. Rev. D \textbf{86}, 044026
(2012).

[7] F S N. Lobo, Phys. Rev D \textbf{71}, 124022 (2005).

[8] A Khaibullina, G Akhtaryanova, R Mingazova, D Saha and R Izmailov, Int.
J. Theor. Phys. \textbf{53}, 1590 (2014).

[9] F S N Lobo, F Parsaei and N Riazi, Phys. Rev. D \textbf{87}, 084030
(2013).

[10] J Katz, D Lynden-Bell and J Bi\v{c}\'{a}k, Phys. Rev. D\textbf{\ 75},
024040 (2007). Erratum: \textit{ibid,} 044901 (2007).

[11] J Katz, D Lynden-Bell and J Bi\v{c}\'{a}k, Class. Quant. Grav. \textbf{%
23}, 9111 (2006).

[12] C Cattoen, T Faber and M Visser, Class. Quant. Grav. \textbf{22}, 4189
(2005).

[13] F S N Lobo, Class. Quant. Grav. \textbf{23}, 1525 (2006).

[14] F S N Lobo and A V B Arellano, Class. Quant. Grav. \textbf{24}, 1069
(2007).

[15] M Visser and D L Wiltshire, Class. Quant. Grav. \textbf{21}, 1135
(2004).

[16] P O Mazur and E Mottola, Proc. Natl. Acad. Sc. USA, \textbf{101}, 9545
(2004).

[17] P Martin-Moruno, N M Garcia, F S N Lobo and M Visser, JCAP 03 (2012)
034.

\end{document}